\documentclass[onecolumn,secnumarabic,amssymb, amsmath, nofootinbib,tightenlines,
nobibnotes, aps, prl,epsfig]{revtex4}
\usepackage{graphicx}% Include figure files
\usepackage{dcolumn}% Align table columns on decimal point
\usepackage{bm}% bold math
\begin{document}
% \preprint{APS/123-QED}
\title{ Investigation of the ratio $\frac{\sigma_{r}}{F_{2}}(Q^2/s,Q^2)$ in the momentum-space approach }% Force line breaks with \\

%\author{S.Fathinejad}
%\altaffiliation{fathinejad2269@gmail.com}%Lines break automatically or can be forced with \\
%\affiliation{Department of Physics, Razi University, Kermanshah
%67149, Iran}% \textbackslash\textbackslash
\author{G.R. Boroun}%
 \email{boroun@razi.ac.ir }
 \affiliation{Department of Physics, Razi University, Kermanshah
67149, Iran}%

\date{\today}% It is always \today, today,
\begin{abstract}
%%%%%%%%%%%%%%%%%%%%%%%%%%%%%%%%%%%%%%%%%%%%%%%%%%%%%%%
We present a calculation of the ratio $\frac{\sigma_{r}}{F_{2}}(x,
Q^2)$ in momentum-space approach using the Block-Durand-Ha (BDH)
parameterization of the proton structure function $F_{2}(x,Q^2)$.
The results are compared with H1 data and extended to high
inelasticity. We also examine the ratio
$\frac{\sigma_{r}}{F_{2}}(\frac{Q^2}{s}, Q^2)$ obtained at a fixed
$\sqrt{s}$ and $Q^2$ to the minimum value of $x$ given by $Q^2/s$,
comparing them with both the HERA data and the color dipole model
bounds. These results and comparisons with HERA data demonstrate
that the suggested method for the ratio $\frac{\sigma_{r}}{F_{2}}$
can be applied in analyses of the Large Hadron Collider and Future
Circular Collider projects. The effect of adding a simple higher
twist term of the form $F_{2}{\ast}H_{2}/Q^2$ to the description
of the ratio $\frac{\sigma_{r}}{F_{2}}(\frac{Q^2}{s}, Q^2)$ at
low-$x$ and low-$Q^2$ values for comparison with the color dipole
bounds and the HERA data is
investigated.\\

%%%%%%%%%%%%%%%%%%%%%%%%%%%%%%%%%%%%%%%%%%%%%%%%%%%%%%%
\end{abstract}
% \pacs{***}%PACS, the Physics and Astronomy
                              %Classification Scheme.
 %Use showkeys class option if keyword
\maketitle
%**********************************************************
%%%%%%%%%%%%%%%%%%%%%%%%%%%%%%%%%%%%%%%%%%%%%%%%%%%%%%%%%%%%%%%%%%%%%%%%%%%%%%%%%%%%%%%%%%%%%
\section{Introduction}

The proton structure functions  from the HERA collider summarize
experimental efforts \cite{H1, H2, H3} conducted by the DESY
collaborations H1 and ZEUS to significantly expand our
understanding of the quark-gluon structure of the proton. These
efforts reveal a sharp increase in $F_{2}(x,Q^2)$ at low values of
$x$ for a constant $Q^2$, where pronounced scaling violations are
observed. These violations are attributed to a high gluon density
within the proton. Measurements of the inclusive deep-inelastic
lepton-nucleon scattering (DIS) cross section have been essential
for testing Quantum Chromodynamics (QCD) and provide us with an
understanding of strong interaction dynamics.\\
The reduced cross section of the measured reaction
$e^{+}p{\rightarrow}e^{+}X$ depends on the proton structure
functions as:
\begin{eqnarray}\label{Reduced_eq}
\sigma_{r}(x,Q^2){\equiv}F_{2}(x,Q^2)-\frac{y^2}{1+(1-y)^2}F_{L}(x,Q^2).
\end{eqnarray}
This dependence is on two independent kinematic variables,
specifically chosen to be $x$ and $Q^2$, as well as on the center
of mass energy squared $s$, with the inelasticity variable
$y=Q^2/sx$. The new kinematic region with small values of the
Bjorken variable $x$ corresponds to the high energy (or Regge)
limit of QCD extended to the Large Hadron electron Collider (LHeC)
\cite{LHeC}. At LHeC, measurements are extended to much lower
values of $x$ and high $Q^2$ with
$\sqrt{s}{\simeq}1.3~\mathrm{TeV}$, which is about 4 times the
center-of-mass (COM) energy range of ep collisions at HERA. At low
values of $x$, the DIS structure functions, which follow the
relation $0{\leq}F_{L}{\leq}F_{2}$ due to the positivity of the
cross sections for longitudinally and transversely polarized
photons scattering off protons, are defined solely by the singlet
quark $xf_{s}(x,Q^{2})$ and gluon density $xf_{g}(x,Q^{2})$ as
\begin{eqnarray} \label{SFs_eq}
F_{k}(x,Q^{2})=<e^{2}>\sum_{a=s,g}\bigg{[}B_{k,a}(x){\otimes}xf_{a}(x,Q^{2})
\bigg{]},~~~k=2,L,
\end{eqnarray}
where $<e^{2}>={\sum_{i=1}^{n_{f}} e_{i}^{2}}/{n_{f}}$ is the
average charge squared for ${n_{f}}$ which ${n_{f}}$ denotes the
number of effective massless flavours. The quantities $B_{k,a}(x)$
are the known Wilson coefficient functions and the parton
densities satisfy the renormalization group evolution equations.
The symbol $\otimes$ indicates convolution over the variable $x$
in the usual form, $f(x){\otimes}g(x)=\int_{x}^{1}
\frac{dz}{z}f(z,\alpha_{s})g(x/z)$.\\
The ratio $\sigma_{r}/F_{2}$ is defined by the following form at
large values of the inelasticity
\begin{eqnarray}\label{Ratio_eq}
\frac{\sigma_{r}}{F_{2}}(x,Q^2)=1-\frac{y^2}{1+(1-y)^2}\frac{F_{L}}{F_{2}}(x,Q^2),
\end{eqnarray}
and in most of the kinematic range the relation
\begin{eqnarray}\label{Approxi_eq}
\frac{\sigma_{r}}{F_{2}}(x,Q^2){\approx}1,
\end{eqnarray}
holds to a very good approximation. Recently, the author in
Ref.\cite{Taylor} has provided a definition the DIS structure
functions at fixed $\sqrt{s}$ and $Q^2$ to the minimum value of
$x$ given by $Q^2/s$ with HERA data. We observe that the ratio
$\sigma_{r}/F_{2}$ simplifies to the ratio $F_{L}/F_{2}$ at the
kinematic point $x_{\mathrm{min}}=Q^2/s$ and $Q^2{\ll}M_{z}^{2}$
by the following form
\begin{eqnarray}\label{Ratio_eq}
\frac{\sigma_{r}}{F_{2}}(Q^2/s,Q^2)|_{y=1}=1-\frac{F_{L}}{F_{2}}(Q^2/s,Q^2).
\end{eqnarray}
An interesting method for the DIS structure functions measurable
in deeply inelastic scattering directly without unobservable
parton distribution functions (PDFs) and without the associated
scheme dependence in the momentum-space (MS) approach is presented
in Ref.\cite{Lappi} and extended based on the Laplace
transformation \cite{Martin1, Martin3, Martin4, Martin5,
Martin6} in Refs.\cite{BH1, BH2}.\\
The ratio of the ${F_{L}}/{F_{2}}$ is estimated to be less than
0.27 according to the color dipole model (CDM) bound in
Ref.\cite{Ewerz1, Ewerz2}, and the bound is lower than 0.27 with
the ratio $\simeq$0.22 in realistic dipole-proton cross section in
Ref.\cite{Niedziela}. The ratio of DIS structure functions is
expressed in terms of the longitudinal-to-transverse ratio of the
photo absorption cross sections in the CDM and defined in
Ref.\cite{Kuroda1, Kuroda1R, Kuroda2R, Kuroda3R} by the following
form
\begin{eqnarray}\label{FL2rho_eq}
\frac{F_{L}(x,Q^{2})}{F_{2}(x,Q^{2})}=\frac{1}{1+2\rho(x,Q^{2})},
\end{eqnarray}
where factor $2$ originates from the difference between the
transverse and longitudinal photon wave function and the factor
$\rho(x,Q^{2})$ is related to the transverse-to-longitudinal ratio
of the photoabsorbtion cross sections as
\begin{eqnarray}\label{rhosigma_eq}
\rho(x,Q^{2})=\frac{\sigma_{T}^{\gamma^{*}p}(W^{2},Q^{2})}{2\sigma_{L}^{\gamma^{*}p}(W^{2},Q^{2})}.
\end{eqnarray}
The transverse and longitudinal inelastic cross-sections in the
dipole picture can be factorized in the following form
\begin{eqnarray}\label{Sigma_eq}
\sigma_{L,T}^{\gamma^{*}p}(x,Q^{2})&=&\int dz
d^{2}\mathbf{r}_{\bot}
|\Psi_{\gamma}^{L,T}(\mathbf{r}_{\bot},z(1-z),Q^{2})|^{2}\sigma_{q\overline{q}}(\mathbf{r}_{\bot},z(1-z),W^{2}),
\end{eqnarray}
where $\Psi_{\gamma}^{L,T}$ are the appropriate spin averaged
light-cone wave functions of the photon and
$\sigma_{q\overline{q}}(r,z,W^{2})$ is the dipole cross-section
which it related to the imaginary part of the $(q\overline{q})p$
forward scattering amplitude. The square of the photon wave
function describes the probability for the occurrence of a
$(q\overline{q})$ fluctuation \cite{Nikolaev, Nikolaev1R,
Nikolaev2R}. The ratio of the DIS structure functions based on the
value of $\rho$ predicted to be $1$ or $\frac{4}{3}$, as
investigated in Refs.\cite{Boroun1, Boroun2, Boroun3, Boroun3R,
Boroun4, Boroun4R} can be used for the ratio
${\sigma_{r}}/{F_{2}}$ to be $\frac{2}{3}$ or
$\frac{8}{11}$ respectively.\\
In this paper, we present the ratio ${\sigma_{r}}/{F_{2}}$ in a
momentum-space approach using the proton structure function
measurable in deeply inelastic scattering based on the
Block-Durand-Ha (BDH) parameterization \cite{Martin1}, which
applies to large and small $Q^2$ and small $x$. The BDH
parameterization provides a better fit to experimental data,
particularly at low values of the Bjorken variable $x$. This
improved fit is crucial for accurately describing the behavior of
the proton structure function in regions where data is sparse.
Additionally, the BDH parameterization aligns well with
theoretical predictions, such as the Froissart bound, which
describes the asymptotic behavior of hadron-hadron cross sections.
By avoiding the need for a specific factorization scheme, the BDH
parameterization simplifies the theoretical calculations involved
in deep inelastic scattering processes, making them more efficient
and accessible.\\
Firstly, we examine the ratio $\frac{\sigma_{r}}{F_{2}}(x,Q^2)$
due to the HERA center-of-mass (COM) energies, then we will extend
the ratio to the limit $x_{\mathrm{min}}=Q^2/s$ which
corresponds to high inelasticity $y=1$, for comparison with
the color dipole model bounds. By including an additional
higher-twist term in the description of the ratio
$\frac{\sigma_{r}}{F_{2}}(Q^2/s,Q^2)$, the results will improve
at low-$x$ and low-$Q^2$ values in comparison with the HERA data.
 The main goal of the paper is the
ratio $\frac{\sigma_{r}}{F_{2}}(Q^2/s,Q^2)$  with the COM energies
at the LHeC and EIC \cite{EIC1, EIC2} colliders. This ratio will
define a limit bound for the
data in these colliders.\\

\section{Method}

In this paper, the ratio ${\sigma_{r}}/{F_{2}}$ is determined
directly in terms of observable quantities, specifically the
proton structure function in DIS utilizing the BDH
parameterization. The structure functions $F_{2}$ and $F_{L}$ for
deeply inelastic scattering into the Parton Distribution Functions
(PDFs) are defined in Ref.\cite{Lappi} by the following forms
\begin{eqnarray} \label{Lappi1_eq}
F_{2}(x,Q^2)&=&<e^2>\bigg{\{}
C^{(0)}_{2,s}+\frac{\alpha_{s}(\mu_{r}^{2})}{2\pi}\bigg{[}C^{(1)}_{2,s}
-\ln{\bigg{(}}\frac{\mu_{r}^{2}}{Q^2}{\bigg{)}}C^{(0)}_{2,s}{\otimes}P_{qq}
\bigg{]}\bigg{\}}{\otimes}x\Sigma(x,\mu_{r}^{2})\nonumber\\
&&+2\sum_{i=1}^{n_{f}}e_{i}^{2}\frac{\alpha_{s}(\mu_{r}^{2})}{2\pi}\bigg{[}C^{(1)}_{2,g}
-\ln{\bigg{(}}\frac{\mu_{r}^{2}}{Q^2}{\bigg{)}}C^{(0)}_{2,g}{\otimes}P_{qg}
\bigg{]}{\otimes}xg(x,\mu_{r}^{2}),
\end{eqnarray}
and
\begin{eqnarray}\label{Lappi2_eq}
F_{L}(x,Q^2)&=&<e^2>\frac{\alpha_{s}(\mu_{r}^{2})}{2\pi}\bigg{\{}
C^{(1)}_{L,s}+\frac{\alpha_{s}(\mu_{r}^{2})}{2\pi}\bigg{[}C^{(2)}_{L,s}
-\ln{\bigg{(}}\frac{\mu_{r}^{2}}{Q^2}{\bigg{)}}C^{(1)}_{L,s}{\otimes}P_{qq}
-2n_{f}\ln{\bigg{(}}\frac{\mu_{r}^{2}}{Q^2}{\bigg{)}}C^{(1)}_{L,g}{\otimes}P_{gq}\bigg{]}\bigg{\}}{\otimes}x\Sigma(x,\mu_{r}^{2})\nonumber\\
&&+2\sum_{i=1}^{n_{f}}e_{i}^{2}\frac{\alpha_{s}(\mu_{r}^{2})}{2\pi}\bigg{\{}C^{(1)}_{L,g}+
\frac{\alpha_{s}(\mu_{r}^{2})}{2\pi}\bigg{[}C^{(2)}_{L,g}-\ln{\bigg{(}}\frac{\mu_{r}^{2}}{Q^2}{\bigg{)}}C^{(1)}_{L,s}{\otimes}P_{qg}
-\ln{\bigg{(}}\frac{\mu_{r}^{2}}{Q^2}{\bigg{)}}C^{(1)}_{L,g}{\otimes}P_{gg}
\bigg{]}\bigg{\}}{\otimes}xg(x,\mu_{r}^{2})\nonumber\\
&&+<e^2>\bigg{(}\frac{\alpha_{s}(\mu_{r}^{2})}{2\pi}\bigg{)}^2\bigg{[}b_{0}\ln{\bigg{(}}\frac{\mu_{r}^{2}}{Q^2}{\bigg{)}}
\bigg{]}\bigg{[}C^{(1)}_{L,s}{\otimes}x\Sigma(x,\mu_{r}^{2})+
2n_{f}C^{(1)}_{L,g}{\otimes}xg(x,\mu_{r}^{2}) \bigg{]},
\end{eqnarray}
where $C_{ij} (i=2,L; j=s,g)$ denotes the scheme-dependent Wilson
coefficient functions. The quark singlet and gluon PDFs can be
expressed in terms of the DIS structure functions, therefore one
of the explicit forms of the evolution equation is reported by the
following form \cite{Lappi}:
\begin{eqnarray}\label{Lappi3_eq}
\frac{dF_{2}(x,Q^2)}{d{\ln}Q^2}&=&\frac{\alpha_{s}(Q^2)}{2\pi}x\bigg{\{}
\frac{1}{4}\bigg{(}\frac{2}{x}-\frac{d}{dx}\bigg{)}\frac{2\pi}{\alpha_{s}(Q^2)}F_{L}(x,Q^2)
+\frac{1}{2}\int_{x}^{1}\frac{dz}{z^2}\frac{2\pi}{\alpha_{s}(Q^2)}F_{L}(z,Q^2)\nonumber\\
&&+C_{F}\bigg{[}\frac{1}{x}F_{2}(x,Q^2)-2\int_{x}^{1}\frac{dz}{z^2}F_{2}(z,Q^2)
+\frac{2}{x}\int_{x}^{1}dz\frac{1}{(1-z)_{+}}F_{2}(\frac{x}{z},Q^2)\bigg{]}\bigg{\}},
\end{eqnarray}
with the color factor $C_{F}=4/3$ associated with the color group
SU(3). The plus function is defined as
\begin{eqnarray}\label{Plus eq}
\int_{x}^{1}dz\frac{f(z)}{(1-z)_{+}}=\int_{x}^{1}dz
\frac{f(z)-f(1)}{1-z}+f(1)\ln(1-x).
\end{eqnarray}
In Ref.\cite{BH1}, the authors developed a method to obtain the
longitudinal structure function, $F^{BH}_{L}$, in the proton
structure function and its derivative using a Laplace-transform
method detailed in \cite{Martin1, Martin3, Martin4, Martin5,
Martin6}. Then in Ref.\cite{BH2}, the authors modified the
equation to significantly improve convergence for increasing
numbers of terms in the series.  Therefore, the ratio
${\sigma_{r}}/{F_{2}}$ is found to be
\begin{eqnarray} \label{Ratio1 eq}
\frac{\sigma_{r}}{F_{2}}(x,Q^2)&=&1-\frac{1}{F^{BDH}_{2}(x,Q^2)}\bigg{\{}4\int_{x}^{1}\frac{d{F}^{BDH}_{2}(z,Q^2)}{d{\ln}Q^2}(\frac{x}{z})^{3/2}\bigg{[}\cos{\bigg{(}}\frac{\sqrt{7}}{2}{\ln}\frac{z}{x}{\bigg{)}}-\frac{\sqrt{7}}{7}
\sin{\bigg{(}}\frac{\sqrt{7}}{2}{\ln}\frac{z}{x}{\bigg{)}}\bigg{]}\frac{dz}{z}-4C_{F}\frac{\alpha_{s}(Q^2)}{2\pi}\nonumber\\
&&{\times}\int_{x}^{1}F^{BDH}_{2}(z,Q^2)(\frac{x}{z})^{3/2}\bigg{[}(1.6817+2\psi(1))\cos{\bigg{(}}\frac{\sqrt{7}}{2}{\ln}\frac{z}{x}{\bigg{)}}
+(2.9542-2\frac{\sqrt{7}}{7}\psi(1))
\sin{\bigg{(}}\frac{\sqrt{7}}{2}{\ln}\frac{z}{x}{\bigg{)}}\bigg{]}\frac{dz}{z}\nonumber\\
&&+8C_{F}\frac{\alpha_{s}(Q^2)}{2\pi}
\bigg{[}\sum_{m=1}^{\infty}\bigg{(}\frac{2(m-4)}{m(m^2-3m+4)}
-\frac{2}{m^2} \bigg{)}\int_{x}^{1}F_{2}^{BDH}(z,Q^2)(\frac{x}{z})^{m}\frac{dz}{z}\nonumber\\
&& + \int_{x}^{1}F_{2}^{BDH}(z,Q^2)\bigg{(}{\rm Li}_ 2
(\frac{x}{z}) -\ln(1-\frac{x}{z})\bigg{)} \frac{dz}{z}
\bigg{]}\bigg{\}},
\end{eqnarray}
where the maximum value of $m$ in the series is chosen to be
approximately $50$, with an accuracy of $1/10^{4}$, which is
sufficient for the present purpose (please refer to the Appendix
in Ref.\cite{BH2}). The structure function $F^{BDH}_{2}(x,Q^2)$
has the following explicit expression:
\begin{eqnarray} \label{BDH eq}
F^{\mathrm{BDH}}_{ 2}(x,Q^{2})=
D(Q^{2})(1-x)^{n}\sum_{m=0}^{2}A_{m}(Q^{2})L^{m},
\end{eqnarray}
with
\begin{eqnarray} \label{Coeff eq}
D(Q^{2})&=&\frac{Q^{2}(Q^{2}+{\lambda}M^{2})}{(Q^{2}+M^{2})^2}, ~
A_{0}=a_{00}+a_{01}L_{2},
 ~A_{i}(Q^{2})=\sum_{k=0}^{2}a_{ik}L_{2}^{k},~ (i=1,2),\nonumber\\
& &L=\ln(1/x)+L_{1},~ L_{1}={\ln}\frac{Q^{2}}{Q^{2}+\mu^{2}},~ L_{2}={\ln}\frac{Q^{2}+\mu^{2}}{\mu^{2}},\nonumber\\
\end{eqnarray}
 where the effective parameters are summarized in
Ref.\cite{Martin1} and are given in Table I.\\
In the limit where $x_{\mathrm{min}}=Q^2/s$ and it is indicated
that the longitudinal polarization of the virtual photon at $y=1$
is zero, we can conclude that the ratio
$\frac{\sigma_{r}}{F_{2}}(Q^2/s,Q^2)|_{y=1}{\rightarrow}1$.
Therefore, we find that this ratio is defined by the following
form:
\begin{eqnarray} \label{Ratio2 eq}
\frac{\sigma_{r}}{F_{2}}(Q^2/s,Q^2)&=&1-\frac{1}{F^{BDH}_{2}(Q^2/s,Q^2)}\bigg{\{}4\int_{Q^2/s}^{1}\frac{d{F}^{BDH}_{2}(z,Q^2)}{d{\ln}Q^2}(\frac{Q^2}{sz})^{3/2}
\bigg{[}\cos{\bigg{(}}\frac{\sqrt{7}}{2}{\ln}\frac{sz}{Q^2}{\bigg{)}}-\frac{\sqrt{7}}{7}\sin{\bigg{(}}\frac{\sqrt{7}}{2}{\ln}\frac{sz}{Q^2}{\bigg{)}}\bigg{]}\frac{dz}{z}\nonumber\\
&&-4C_{F}\frac{\alpha_{s}(Q^2)}{2\pi}\int_{Q^2/s}^{1}F^{BDH}_{2}(z,Q^2)(\frac{Q^2}{sz})^{3/2}\bigg{[}(1.6817+2\psi(1))\cos{\bigg{(}}\frac{\sqrt{7}}{2}{\ln}\frac{sz}{Q^2}{\bigg{)}}\nonumber\\
&&+(2.9542-2\frac{\sqrt{7}}{7}\psi(1))\sin{\bigg{(}}\frac{\sqrt{7}}{2}{\ln}\frac{sz}{Q^2}{\bigg{)}}\bigg{]}\frac{dz}{z}
+8C_{F}\frac{\alpha_{s}(Q^2)}{2\pi}
\bigg{[}\sum_{m=1}^{\infty}\bigg{(}\frac{2(m-4)}{m(m^2-3m+4)}
-\frac{2}{m^2}
\bigg{)}\nonumber\\
&&{\times}\int_{Q^2/s}^{1}F_{2}^{BDH}(z,Q^2)(\frac{Q^2}{sz})^{m}\frac{dz}{z}
 + \int_{Q^2/s}^{1}F_{2}^{BDH}(z,Q^2)\bigg{(}{\rm Li}_ 2
(\frac{Q^2}{sz}) -\ln(1-\frac{Q^2}{sz})\bigg{)} \frac{dz}{z}
\bigg{]}\bigg{\}}.
\end{eqnarray}
In the following section, we calculated the ratio of
$\frac{\sigma_{r}}{F_{2}}$ over a wide range of inelasticity
based on the H1 data.\\

\section{Results and Conclusion}

With the explicit form of the ratio $\frac{\sigma_{r}}{F_{2}}$
(i.e., Eq.~(\ref{Ratio1 eq})), we begin to extract the numerical
results at small $x$ in a wide range of the inelasticity $y$,
using the parametrization of $F^{BDH}_{2}(x,Q^2)$ (i.e.,
Eq.~(\ref{BDH eq})). The QCD parameter $\Lambda$ for four numbers
of active flavor has been extracted \cite{Lipatov} due to
$\alpha_{s}(M_{z}^{2})=0.1166$ with respect to the LO form of
$\alpha_{s}(Q^2)$ with $\Lambda=136.8~\mathrm{MeV}$. In order to
account for the effect of the production threshold for the charm
quark
 with $m_{c}=1.29^{+0.077}_{-0.053}~\mathrm{GeV}$ \cite{H4, H5}, the
rescaling variable $\chi$ is defined by the form
$\chi=x(1+4\frac{m_{c}^{2}}{Q^2})$ which is reduced to the Bjorken
variable $x$ at high $Q^2$.\\
In Fig.1, we show the ratio of $\frac{\sigma_{r}}{F_{2}}$ based on
the $F_{2}$ and $F_{L}$ parametrizations in \cite{Martin1} and
\cite{BH1, BH2}, respectively. The comparison of the ratio with
the H1 data \cite{H1} at moderate and low inelasticity is
excellent according to the uncertainties. The ratio at large $x$
(low inelasticity) is approaching unity. The ratio decreases from
unity as inelasticity increases ( very low $x$). This behavior is
more pronounced at low $Q^2$ values. In Fig.1, we depict this
behavior up to $y=1$ as experimental data have not been determined
in this region. The H1 data are selected in the region
$1.5{\leq}Q^2{\leq}45~\mathrm{GeV}^2$ at the interval
$0.518{\leq}y{\leq}0.004$. The importance of the longitudinal
structure functions is evident in experimental data at low $Q^2$
values and high inelasticity, where the photon interaction becomes
hadron-like in the CDM, as discussed in Ref. \cite{Boroun2}. The
error bars for the ratio $\frac{\sigma_{r}}{F_{2}}$ are determined
by the following formula:
\begin{eqnarray} \label{Error1 eq}
 \Delta(\frac{\sigma_{r}}{F_{2}})=\frac{\sigma_{r}}{F_{2}}\sqrt{\bigg{(}\frac{\Delta{\sigma_{r}}}{F_{L}}\bigg{)}^2
 +\bigg{(}\frac{\Delta{F_{2}}}{F_{2}}\bigg{)}^2},
 \end{eqnarray}
where in the H1 data, $\Delta{\sigma_{r}}$ and  $\Delta{F_{2}}$
are obtained from the H1 experimental data \cite{H1}. In our
calculations, the error bands depend on the uncertainties of
$F_{2}(x,Q^2)$ and $F_{L}(x,Q^2)$ according to the parametrization
coefficients in the BDH model (Table I) as follows:
\begin{eqnarray} \label{Error2 eq}
 \Delta(\frac{\sigma_{r}}{F_{2}})=\frac{y^2}{Y_{+}}\frac{F_{L}}{F_{2}}\sqrt{\bigg{(}\frac{\Delta{F_{L}}}{F_{L}}\bigg{)}^2
 +\bigg{(}\frac{\Delta{F_{2}}}{F_{2}}\bigg{)}^2}.
 \end{eqnarray}
\begin{figure}[h]
\includegraphics[width=0.8\textwidth]{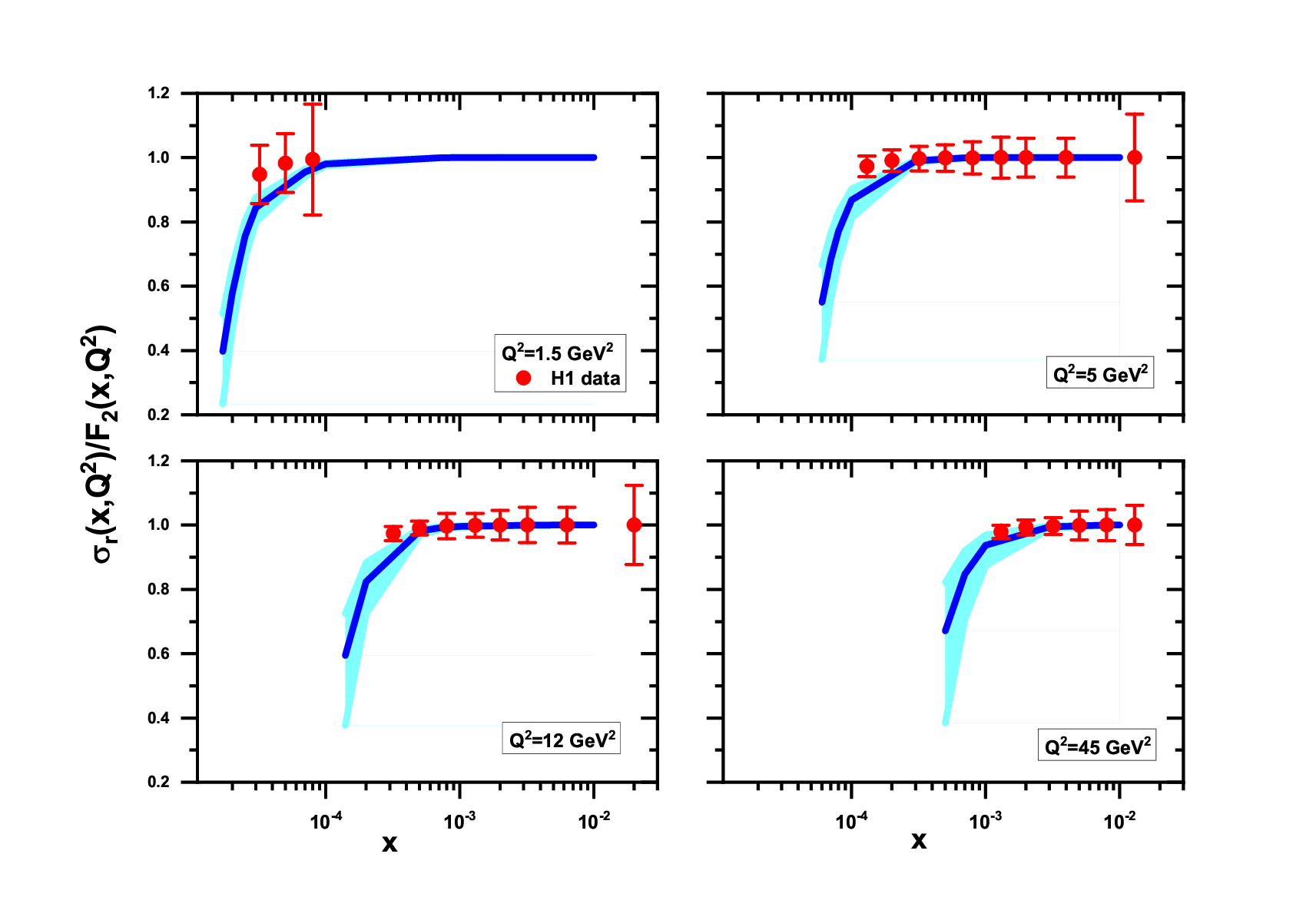}
\caption{The extracted ratio $\frac{\sigma_{r}}{F_{2}}$ (blue
curves) from the parametrization methods
 is compared with the H1 data (red circles) \cite{H1}. The total
 errors, which account for both the reduced cross sections and the
 structure functions,
  are included. The error  bands (cyan curves) of
the ratio $\frac{\sigma_{r}}{F_{2}}$ correspond to the uncertainty
in the parameterization of $F_{2}$ and $F_{L}$ as shown in Table
I.}\label{Fig1}
\end{figure}
\begin{figure}[h]
\includegraphics[width=0.6\textwidth]{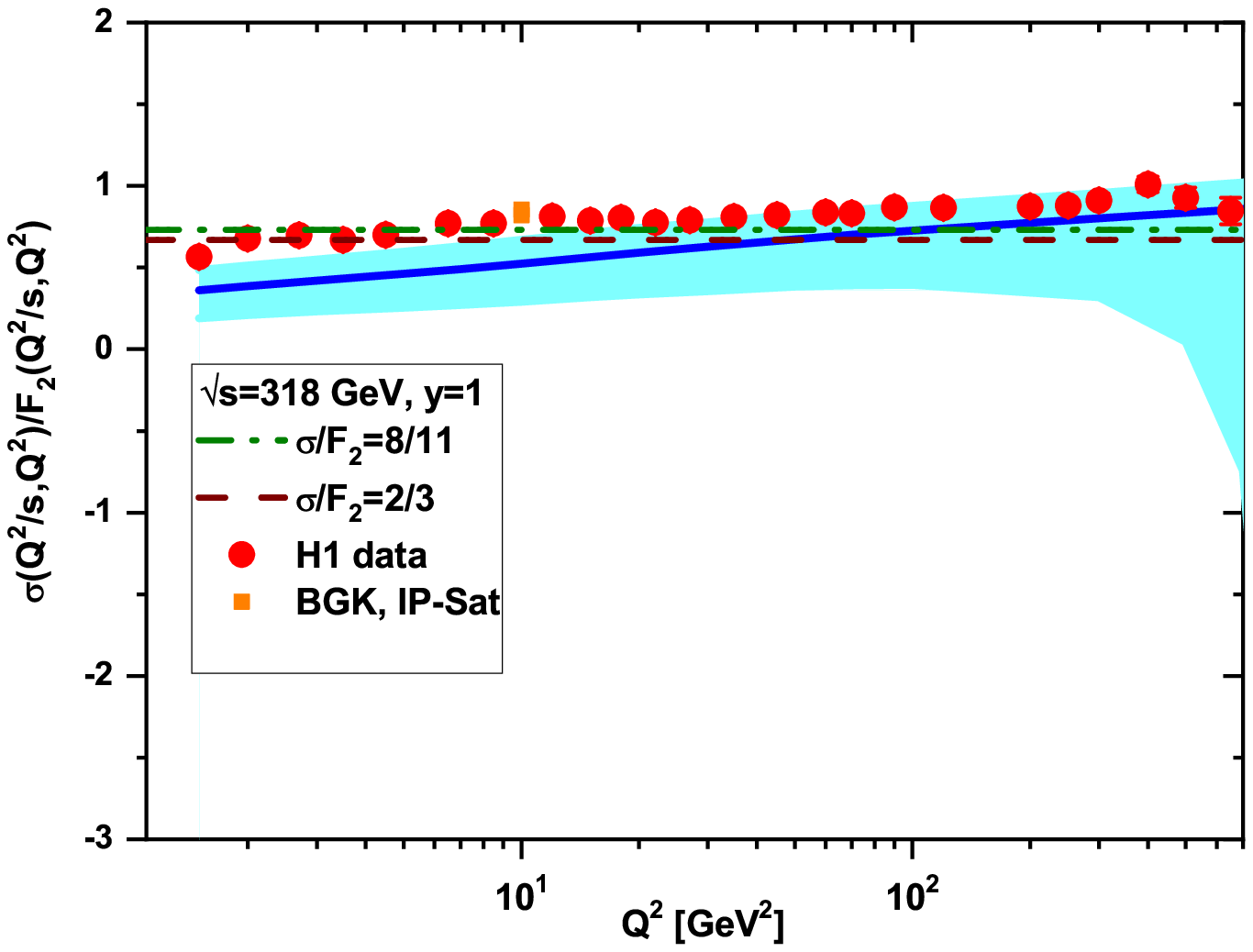}
\caption{ We plot the ratio
$\frac{\sigma_{r}}{F_{2}}(\frac{Q^2}{s}, Q^2)$ as  a function of
$Q^2$ at $y=1$ for the HERA NC ep inclusive scattering data with
$\sqrt{s}=318~\mathrm{GeV}$. The blue curves are extracted and
compared with the results in Table VIII of Ref.\cite{Taylor} (red
circles), as well as the BGK and IP-Sat models (yellow squares).
The error
 bands correspond to the uncertainty in the
parameterization of $F_{2}$ in \cite{Martin1}. The dipole upper
bounds are represented by dashed and dashed-dot lines
corresponding to $\rho=1$ and $\frac{4}{3}$ in the CDM.
}\label{Fig2}
\end{figure}
In Fig. 2, we present the prediction of Eq.~(\ref{Ratio2 eq}) for
the ratio $\frac{\sigma_{r}}{F_{2}}(\frac{Q^2}{s}, Q^2)$ and
compare it with the results reported in Ref.\cite{Taylor} at
$\sqrt{s}=318~\mathrm{GeV}$  for $y=1$, along with the data
uncertainties. As shown in this figure, the values of the ratio
are comparable to the data from Ref.\cite{Taylor} and are in good
agreement with the CDM bounds over a wide range of $Q^2$ values.
We also compared the ratio with the results proposed by the
authors in Ref.\cite{Machado} within the CDM in light of HERA
high-precision data based on the impact parameter saturation model
(IP-Sat) \cite{Kowalski} and the BGK model \cite{Bartels}.
According to the predictions of the DIS structure functions of
models BGK and IP-Sat, there are only two data points with high
inelasticity at $\sqrt{s}=318~\mathrm{GeV}$ in Table IV of
Ref.\cite{Machado}. These results demonstrate good agreement
between models and the MS approach.\\
Indeed, Fig.2  compares  the HERA data in the phase
space region with the color dipole approach, showing that the success of
the CDM in this kinematic region is due to the all twist
resummation embedded within it. Higher twist (HT) corrections in
deeply inelastic scattering within the saturation model are
defined in the literature \cite{HT1, HT2}. A twist analysis of the
nucleon structure functions $F_{T}$ and $F_{L}$ at small values of
the Bjorken variable $x$ reveals that for $F_{L}$,
the higher twist corrections are significant, while for $F_{2}=F_{T
}+F_{L}$ there is nearly complete cancellation of twist-4
corrections in $F_{T}$ and $F_{L}$. A twist decomposition of the
proton structure functions $F_{2}$ and $F_{L}$ obtained from the
Balitsky-Kovchegov (BK) equation using the Mellin representation
of the scattering cross-sections at high energies based on the effects
of non-linear small-$x$ evolution of the gluon distribution within
the collinear approximation framework is performed in
Refs.\cite{HT3, HT4}.\\
\begin{figure}[h]
\includegraphics[width=0.6\textwidth]{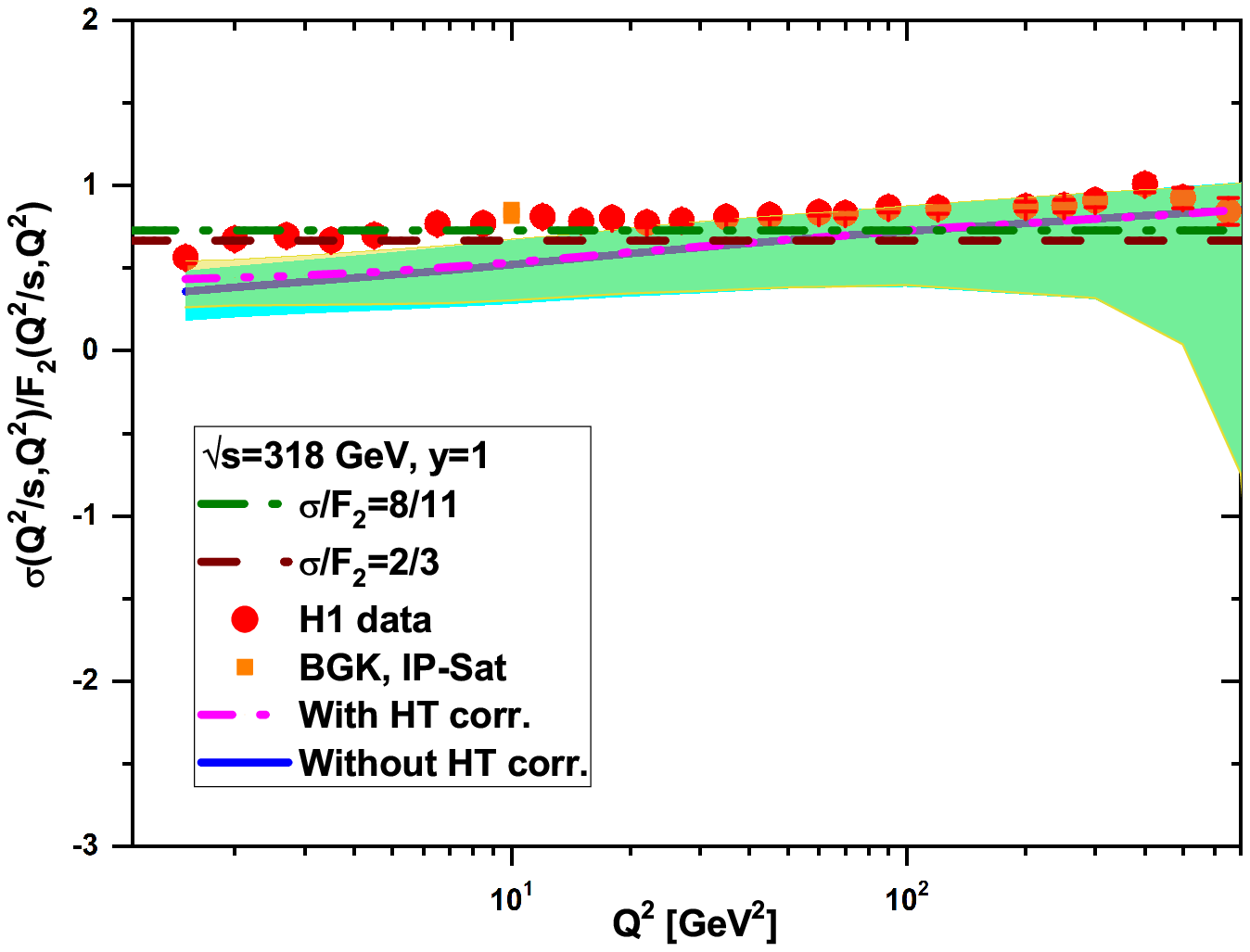}
\caption{ We plot the ratio
$\frac{\sigma_{r}}{F_{2}}(\frac{Q^2}{s}, Q^2)$ as  a function of
$Q^2$ at $y=1$ for the HERA NC ep inclusive scattering data with
$\sqrt{s}=318~\mathrm{GeV}$ with (magenta-solid curve) and without
(blue-solid curve) HT corrections. The results are compared with
the data in Table VIII of Ref.\cite{Taylor} (red circles), as well
as the BGK and IP-Sat models (yellow squares). The error
 bands correspond to the uncertainty in the
parameterization of $F_{2}$ in \cite{Martin1} and the HT
coefficient in \cite{HT5, HT6, HT7}. The dipole upper bounds are
represented by dashed and dashed-dot lines corresponding to
$\rho=1$ and $\frac{4}{3}$ in the CDM. }\label{Fig3}
\end{figure}
In Fig.3, we consider HT corrections as multiplicative shifts to
the DIS structure function $F_{2}$, represented by the ratio
$\frac{\sigma_{r}}{F_{2}}(\frac{Q^2}{s}, Q^2)$ in Eq.~(\ref{Ratio2
eq}), which  depends on ${F_{2}}(\frac{Q^2}{s}, Q^2)$. These HT
corrections are parameterized as a phenomenological
unknown function, and the values of the unknown parameters are
determined through fits to experimental data \cite{HT5, HT6, HT7}.
It is common practice to adjust the leading-twist structure function by
incorporating a term inversely proportional to $Q^2$ as a
phenomenological power correction to account for HT effects in the structure function:
\begin{eqnarray} \label{HT eq}
F_{2}(x,Q^2)=F_{2}^{LT}(x,Q^2)\left(1+\frac{H_{2}(x)}{Q^2}
\right).
 \end{eqnarray}
Here,  $F_{2}^{LT}$ represents the leading twist contribution to $F_{2}$, and the HT coefficient function
$H_{2}(x)$ is parameterized as:
\begin{eqnarray} \label{H2x eq}
H_{2}(x)=\sum h_{\alpha}f_{\alpha}(x).
 \end{eqnarray}
The parameters in this equation capture the HT effects in
perturbative QCD, as discussed in Refs.\cite{HT8, HT9}. In the
studies by authors in Refs.\cite{HT5, HT6, HT7},  the inclusion of
an HT term in $F_{2}$ at low-$x$ and low-$Q^2$ was investigated,
revealing the necessity for $H_{2}=0.12{\pm}0.07~\mathrm{GeV}^2$
for  data with $y<1$. In this figure (i.e., Fig.3), the ratio
$\frac{\sigma_{r}}{F_{2}}(\frac{Q^2}{s}, Q^2)$ is plotted against
$Q^2$ values at $x_{\mathrm{min}}=Q^2/s$ with and without HT
corrections, compared  with to results from Ref.\cite{Taylor} at
$\sqrt{s}=318~\mathrm{GeV}$  for $y=1$, accompanied by data
uncertainties, and  CDM bounds. The impact of HT contribution on
the ratio $\sigma_{r}/F_{2}$
is noticeable at low $Q^2$ values.\\
The HT coefficient of $H_{2}$ is determined through a fit to data
over a wide range with $y<1$. To demonstrate the HT effects in the
ratio $\frac{\sigma_{r}}{F_{2}}(\frac{Q^2}{s}, Q^2)$ at $y=1$, we
calculated the ratio across a board range of the HT coefficient.
In Fig.4, we present the HT effects with $H_{2}=0.12, 0.5$, and
$1$ in comparison to the HERA data.
\begin{figure}[h]
\includegraphics[width=0.6\textwidth]{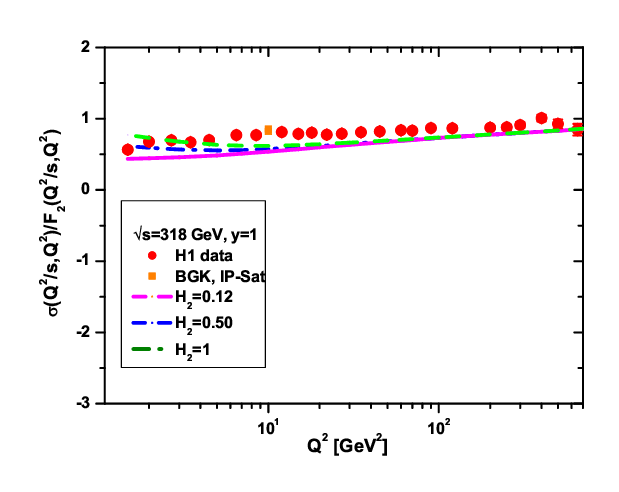}
\caption{ We plot the ratio
$\frac{\sigma_{r}}{F_{2}}(\frac{Q^2}{s}, Q^2)$ as a function of
$Q^2$ at $y=1$ for the HERA NC ep inclusive scattering data with
$\sqrt{s}=318~\mathrm{GeV}$, incorporating HT corrections over a wide range of $H_{2}$ coefficients: $H_{2}=0.12$ (magenta-solid curve),
$0.5$ (blue-dashed-dot curve) and $1$ (green-dashed curve).
 These results are compared to
the data in Table VIII of Ref.\cite{Taylor} (red circles), as well
as the BGK and IP-Sat models (yellow squares). }\label{Fig4}
\end{figure}
We observe that the results improve with increasing the HT coefficient
due to high inelasticity, compared to the fit
parameter $H_{2}=0.12$ at $y<1$. The origin of this discrepancy
is model-dependent, with all twist resummation improving the model as
discussed in Refs.\cite{HT3, HT4}.\\
Future measurements at the EIC and LHeC will be examined at
upcoming colliders,  serving as an interesting tool to
investigate the longitudinal structure function at low $x$ and
moderate $Q^2$. Our predictions for the ratio
$\frac{\sigma_{r}}{F_{2}}(\frac{Q^2}{s}, Q^2)$ in the kinematic
range of the EIC and  LHeC colliders are shown in Fig.5. In this
figure, the behavior of the ratio
$\frac{\sigma_{r}}{F_{2}}(\frac{Q^2}{s}, Q^2)$ is considered with COM
energies $\sqrt{s}=89~\mathrm{GeV}$ and $1.3~\mathrm{TeV}$ for the
EIC and LHeC colliders respectively, and compared
with the CDM bounds. The uncertainties in the ratio are due to the
parametrization of the proton structure function and dependence on
the model. For the EIC COM energy,  results at
$Q^2>100~\mathrm{GeV}^2$ are out of reach due to large uncertainties in that domain. These results are independent of the
mass number A and depend on the COM energies.\\
\begin{figure}[h]
\includegraphics[width=0.6\textwidth]{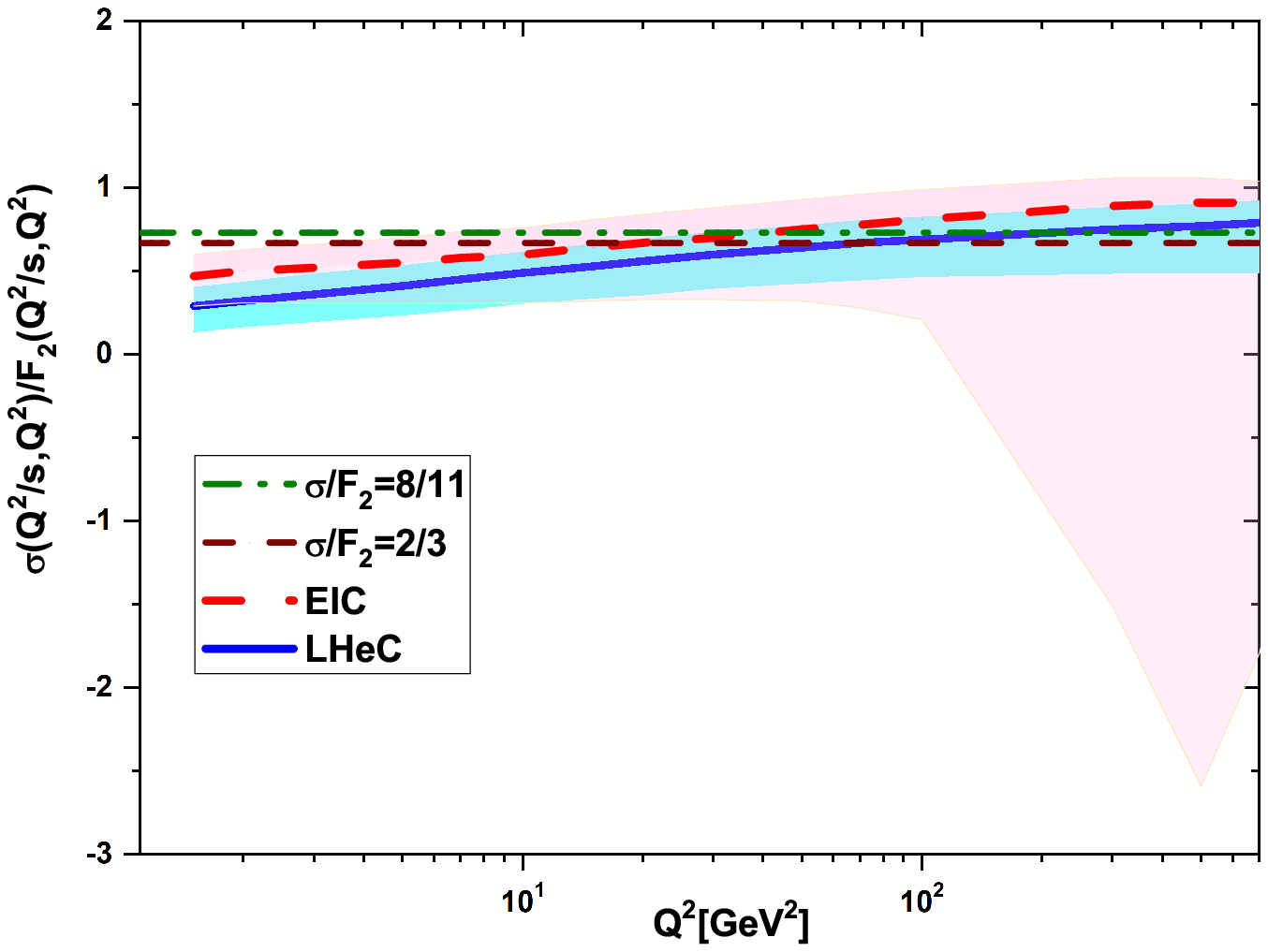}
\caption{ We plot the ratio
$\frac{\sigma_{r}}{F_{2}}(\frac{Q^2}{s}, Q^2)$ as  a function of
$Q^2$ at $y=1$ for the EIC (red-dashed curve) and the LHeC
(blue-solid curve) COM energies with $\sqrt{s}=89~\mathrm{GeV}$
and $1.3~\mathrm{TeV}$ respectively. The error
 bands correspond to the uncertainty in the
parametrization of $F_{2}$ in \cite{Martin1}. The dipole upper
bounds are represented by dashed and dashed-dot lines
corresponding to $\rho=1$ and $\frac{4}{3}$ in the CDM.
}\label{Fig5}
\end{figure}

In conclusion, we studied the ratio $\frac{\sigma_{r}}{F_{2}}(x,
Q^2)$ in momentum space using the Block-Durand-Ha parameterization
of the proton structure function $F_2(x, Q^2)$. We developed a
method to analyze high inelasticity, where
$x_{\mathrm{Bj}}=x_{\mathrm{min}}=Q^2/s$. The extraction procedure
was explained for analyzing the ratio of
$\frac{\sigma_{r}}{F_{2}}(\frac{Q^2}{s}, Q^2)$ in the kinematical
region of the H1 collaboration data and extended to high
inelasticity data, which is determined by the
 effective parameters of the BDH parameterization and compared with
 the CDP bounds.\\
Our study further examines the ratio
$\frac{\sigma_{r}}{F_{2}}(\frac{Q^2}{s}, Q^2)$, comparing it to H1
data and bounds from the color dipole model, showing strong
agreement that supports the proposed methods. Additionally, we
compared the results with the predictions of the BGK and IP-Sat
models, finding good agreement with datasets across a wide
range of $Q^2$ values.\\
The HT corrections to the proton structure function by adding a
simple form $F_{2}{\ast}H_{2}/Q^2$ to  the ratio
$\frac{\sigma_{r}}{F_{2}}(\frac{Q^2}{s}, Q^2)$ at low-$x$ and low-
$Q^2$ values are considered. These effects show that the ratio
increases as the $Q^2$ value decreases at high inelasticity. The
ratio $\frac{\sigma_{r}}{F_{2}}(\frac{Q^2}{s}, Q^2)$ at the EIC
and the LHeC COM energies are considered in a wide range of $Q^2$
and compared with the CDM bounds.\\

%%%%%%%%%%%%%%%%%%%%%%%%%%%%%%%%%%%%%%%%%%%%%%%%%%%%%%%%%%%
\section{ACKNOWLEDGMENTS}

The author is thankful to  Razi University for the financial
support provided by this project. \\

%%%%%%%%%%%%%%%%%%%%%%%%%%%%%%%%%%%%%%%%%%%%%%%%%%%%%%%%%%%%%%%%%%%%%%%%
\begin{table} [h]
\caption{The effective parameters in the BDH expression for $
F_2(x,Q^2)$  at small $x$ for
$0.15~\mathrm{GeV}^{2}<Q^{2}<3000~\mathrm{GeV}^{2}$ provided by
the following values. The fixed  parameters are defined by the
Block-Halzen fit to the real photon-proton cross section as
$M^{2}=0.753 \pm 0.068~ \mathrm{GeV}^{2}$, $\mu^2 = 2.82 \pm
0.290~ \mathrm{GeV}^{2}$, and $a_{00}=0.2550\pm 0.016$
\cite{Martin1}.}
\begin{tabular} {cccc}
\toprule \\  \multicolumn{2}{c}{parameters \quad \quad \quad ~~~~~~~~~~~~~~~~value}    \\ &&&\\ \hline \\ &&&\\

  $n$ & \quad  $n=11.49 \pm 0.99$ & &\\

 $\lambda$ & \quad  $2.430~\pm 0.153 $ & &\\

&&&\\

$a_{01}$& \quad  $1.475\times 10^{-1}~\pm 3.025\times10^{-2}$ & &\\

&&&\\

  $a_{10} $  &   \quad  $8.205\times 10^{-4}~~  \pm  4.62\times10^{-4} $  \\

  $a_{11} $  &   \quad   $-5.148\times 10^{-2}\pm 8.19\times10^{-3}$  \\

  $a_{12}$   &    \quad  $-4.725\times 10^{-3}\pm 1.01\times10^{-3}$   \\  &&&\\

 $a_{20}$   &   \quad   $2.217\times 10^{-3}\pm 1.42\times10^{-4} $ \\

 $a_{21}$   &   \quad   $1.244\times 10^{-2}\pm 8.56\times10^{-4}$  \\

 $a_{22}$    &    \quad  $5.958\times 10^{-4}\pm 2.32\times10^{-4} $ \\ &&& \\

\hline

\end{tabular}
\end{table}
%%%%%%%%%%%%%%%

%%%%%%%%%%%%%%%%%%%%%%%%%%%%%%%%%%%%%%%%%%%%%%%%%%%

\end{document}